\documentclass[12pt,dvips]{article}
\usepackage{epsfig}
\begin{document}

\thispagestyle{empty}

\date{\today}
\title{
\vspace{-5.0cm}
\begin{flushright}
{\normalsize UNIGRAZ-}\\
\vspace{-0.3cm}
{\normalsize UTP-}\\
\vspace{-0.3cm}
{\normalsize 26-02-96}\\
\end{flushright}
\vspace*{2.5cm}
Universality of the Ising model on sphere--like lattices}
\author
{\bf Ch. Hoelbling and C.B. Lang \\  \\
Institut f\"ur Theoretische Physik,\\
Universit\"at Graz, A-8010 Graz, AUSTRIA}
\maketitle
\begin{abstract}
We study the 2D Ising model on three different types of lattices that
are topologically equivalent to spheres. The geometrical shapes are
reminiscent of the surface of a pillow, a 3D cube and a sphere,
respectively.  Systems of volumes ranging up to O($10^5$) sites are
simulated and finite size scaling is analyzed.  The partition function
zeros and the values of various cumulants at their respective peak
positions are determined and they agree with the  scaling behavior
expected from universality with the Onsager solution on the torus
($\nu=1$). For the pseudocritical values of the coupling we find
significant anomalies indicating a shift exponent $\neq 1$ for
sphere--like lattice topology.
\end{abstract}

\newpage

\section{Motivation and introduction}

Most of the Monte Carlo studies of spin models have been done on
hypercubic lattices with periodic boundary conditions, i.e.  with the
topology of a torus.  Recently, however, there has been a growing
interest  in the effect of other topologies. Some of that interest is
motivated by the advance of quantum gravity models, some by problems
related to the role of topological excitations of the particular
system.

Various investigations [1--6]
\nocite{LaNe94b,JeLaNe95,DiGoSa94,HoJa95,GaLa95,HoJaJe95} have been
carried out on lattices topologically equivalent to the surface of a
hypersphere in $D+1$ dimensions. On sphere--like surfaces a loop can be
continuously contracted to a point: The fundamental homotopy group is
trivial. This has immediate consequences for the dynamics of
string--like objects like monopole loops (in 4D pure gauge theory with
$U(1)$ gauge symmetry) or the boundaries of clusters in 2D spin
models.  We therefore might expect a different approach to the
thermodynamic limit and different finite size corrections to scaling.

At first order phase transitions the phase mixture obtained in the
thermodynamic limit may be influenced by local changes, like fixing one
spin, or changing the boundary conditions.  The critical exponents
however, should not depend on such effects. At second order phase
transitions one expects that the critical properties are not affected
by boundary conditions or by  whether one works on lattices with torus
or spherical surface geometries if either becomes flat in the infinite
volume limit. Such an assumption of universality should be tested in
practical examples, which is one of the motivations for this work.
Another question is whether the finite size scaling (FSS) ansatz is
general enough to persist. The important leading terms and the
corrections to them may have different size depending on the lattices
and their topology as has been observed in the 4D and 2D studies.

The Ising model is explicitly solved for torus geometry \cite{On44}
even on finite lattices \cite{Ka49,FeFi69}. It has a well understood
$2^{\mbox{\rm \footnotesize nd}}$ order phase transition.  FSS of the
bulk quantities is dominated by the leading term and provides a good
example of the power of this method to determine critical indices.

Here we want to present a Monte Carlo study for the Ising model on
different 2D lattices with the topology of the surface of a sphere: the
surface of a 3D cube, a pillow--like structure and the cubic surface
projected to a sphere. In all these cases we determine cumulants and
partition function zeros on different lattice sizes in order to study
the size dependence and FSS in detail.  After the introduction of these
geometries in sect.\ref{sec2}, the simulation and multi--histogram
analysis are discussed in sect.\ref{sec3} and the results in
sect.\ref{sec4}. We conclude that there are sizable differences to the
usual results on torus--like lattices, but that universality prevails.
Preliminary results of this work have been presented in
\cite{HoJaJe95}.

\section{Lattice geometry\label{sec2}}

The most straightforward approach to constructing a sphere--like lattice
would be triangulation. We want to stay as close as possible to the
original action, however. This is partly motivated by the related
studies \cite{LaNe94b} and \cite{GaLa95}, where one wants to keep a
4-link plaquette structure. Also, two of
the lattices studied here can be viewed at as combinations of the usual
square lattices glued together at the boundaries; thus
in the thermodynamic limit even the non--universal critical coupling
ought to agree with that for torus geometry.

\begin{figure}
\begin{center}
\epsfig{file=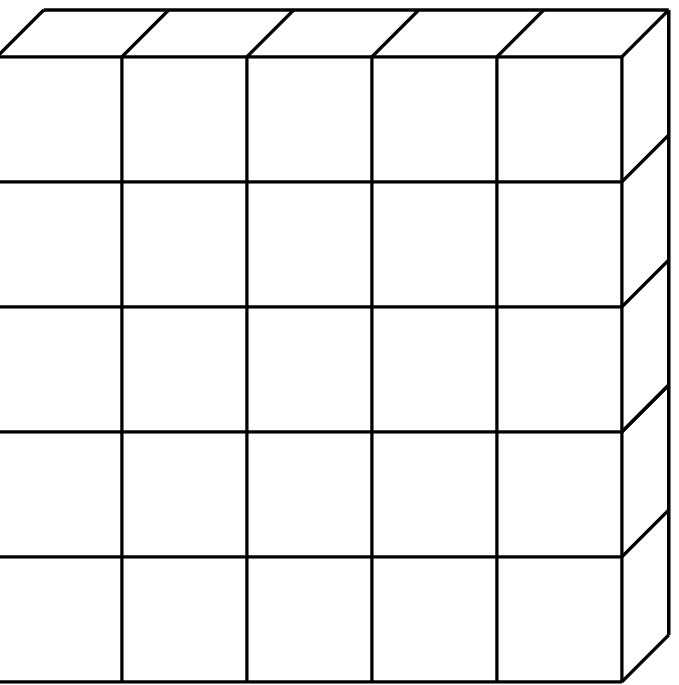,width=3cm}
\hfil
\epsfig{file=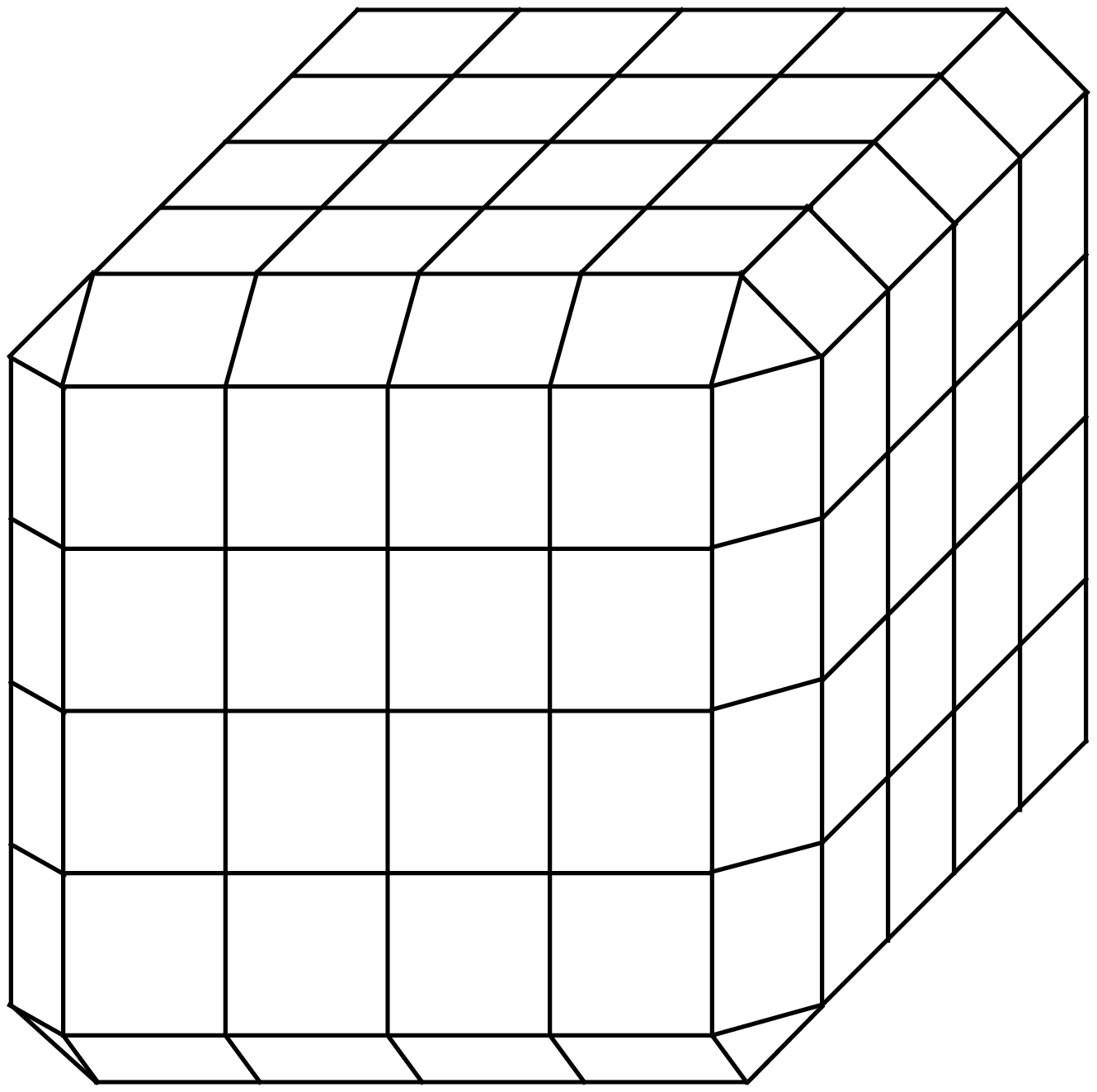,width=3.8cm}
\hfil
\epsfig{file=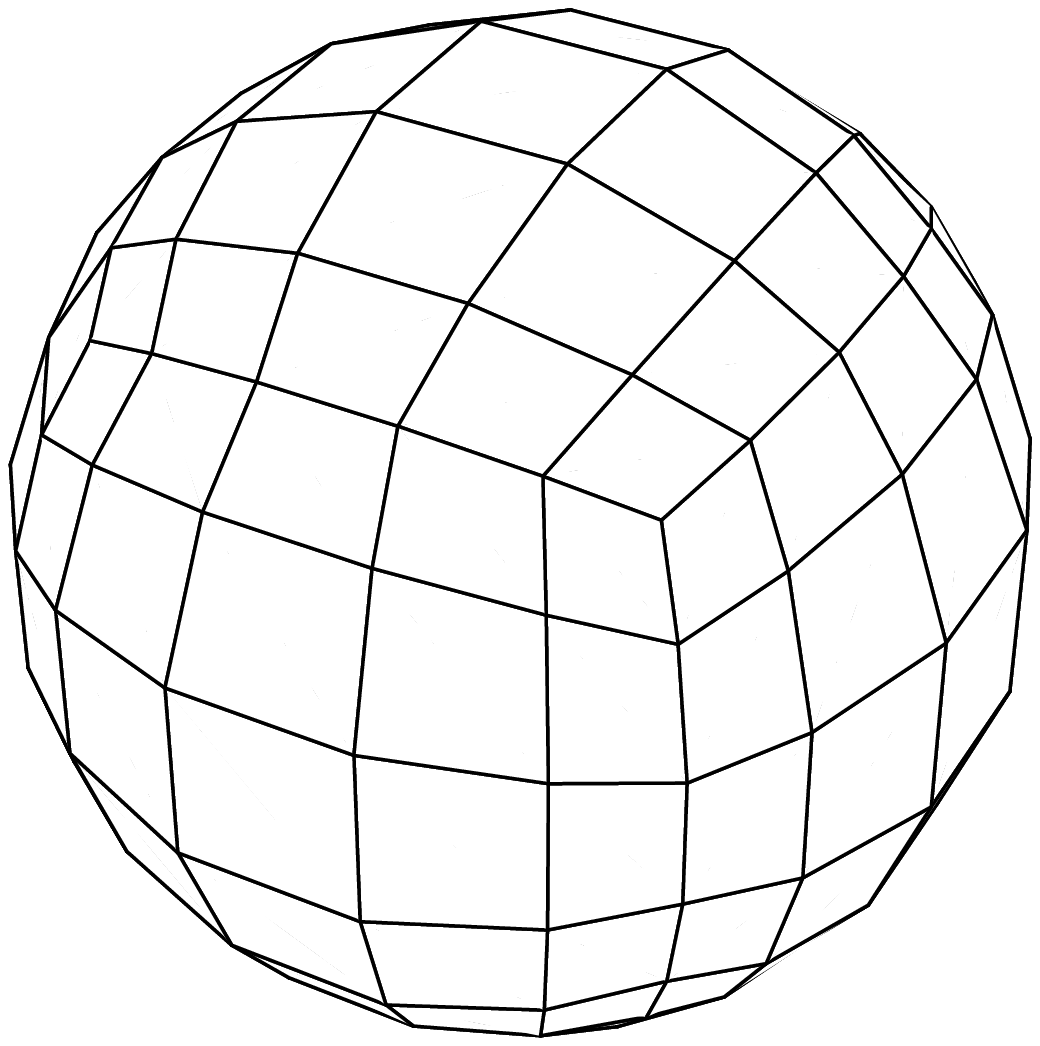,width=4cm}
\end{center}
\caption{Lattices ${P\!I\,}$[6], ${S\!H\,}$[6] and
$S$[6].\label{LattFig}}
\end{figure}

The torus is our reference geometry; in addition three other lattice
geometries were used for our simulations (cf. fig. \ref{LattFig}).
\begin{description}
\item[Torus ${T\!O\,[N]}$:]
The solution for the Ising model on an
$N\times N$ lattice with periodic b.c. is known \cite{Ka49,FeFi69}
and we will use this as the standard to compare our results with.
\item[Pillow ${P\!I\,[N]}$:]
This is the surface of a cube with $N \times N \times 2$ sites, where
$N$ will be called the base length in the subsequent discussion.
Basically it is made out of two $N\times N$ lattices glued together at
the edges. The curvature is concentrated in the $4 \times 2$ corners,
elsewhere the lattice is locally flat.  This lattice ${P\!I\,}$[N] has
$2 N^2$ sites and $4(N^2-1)$ links.
\item[(Dual) Cube ${S\!H\,[N]}$:]
This lattice is dual to the surface of a 3D cubic lattice of base
length $N$. Each plaquette of the cubic surface is identified with a
site of ${S\!H\,}$. This has the advantage, that each site has 4 links
to nearest neighbors, like for the torus. The lattice has $6(N-1)^2$
sites and $12(N-1)^2$ links. The curvature is concentrated on the
$8\times 3$ corner points.
\item[Sphere ${S\,[N]}$:]
That lattice is defined by the sites on the surface of a $N^3$ cube,
projected onto the unit sphere. The site--link connectivity structure
is identical to that of the cube's surface, the only difference lies in
weight factors in the action, to be discussed below.  The lattice has
$6(N-1)^2 + 2$ sites and $12 (N-1)^2$ links.
\end{description}

The total energy (or action, if used in the context of quantum field
theories) for the model is given by
\begin{equation} \label{ActionDef}
E=-\sum_{x,\mu} w_{x,\hat{\mu}}s_x s_{x+\hat{\mu}}
\end{equation}
where the sum runs over all links (at sites $x$ in directions
$\hat\mu$).  The spin variables $s_x$ have values $\in\mathbf{Z}(2)$.
The weight factors $w_{x,\hat{\mu}}$ are equal to 1 for the torus and
the lattice types ${P\!I\,}$ and ${S\!H\,}$.  The
partition function is
determined by the sum
\begin{equation}
Z = \sum_C \exp{(-\beta E(C))}
\end{equation}
over all spin configurations $C$.

For lattices ${P\!I\,}$ and ${S\!H\,}$ the
curvature is concentrated around the corners. The deficit angle denotes
the deviation of the sums of angles of plaquettes (or triangles) at a
given site from the flat--space value $2\pi$. For ${P\!I\,}$
its value is $\pi/2$ on each of the $4\times 2$ corners, for
${S\!H\,}$ it is $\pi/6$ on each of the $8\times 3$
corners.  It vanishes on all other sites, as it does on the torus: The
lattices are flat almost everywhere. The total curvature on the
sphere--like lattices therefore has the value $4\pi$ or Euler number 2,
as compared to 0 for the torus.  Euler's relation for sphere--like
lattices is $n_s+n_p-n_l=2$ ($n_s$, $n_p$, and $n_l$ denote the total
number of sites, plaquettes and links, respectively) whereas for torus
topology this sum is zero.

Both lattice types ${P\!I\,}$ and ${S\!H\,}$ can be
imagined as built out of flat
$N\times N$ pieces, glued together along their boundaries. The
deviation from the torus shaped lattice thus disappears at least as
fast as a boundary contribution $O(1/N)$. The contribution from the
corners, where the curvature is concentrated, is suppressed
$O(1/N^2)$.  We expect that, although not a universal quantity, the
value of the critical coupling $\beta_c$ (in the thermodynamic limit)
coincides with that for the torus.

In order to mimic a truly spherical lattice more closely, we also took
the structure of a cubic surface lattice projected to the sphere as
shown in fig. \ref{LattFig}. Since the link connectivity structure
does not change, the difference has to be expressed by modifying the
weight factors accordingly. Christ et al. \cite{ChFrLe82} discuss a
possible definition for triangulated (random) surfaces, which obeys
certain properties, that are necessary for the consistency of the
continuum field theory at a second order phase transition.

Our lattices are built out of quadrangles and therefore we have to
modify that method for the derivation of the weight factors.  We start
with the scalar continuum field theory with the Euclidean invariant
kinetic term
\begin{equation}
\int d^2x \left(\partial_\mu\phi(x)\right)^2 \quad.
\end{equation}
(In fact, we can define the Ising model as a discretization of a scalar
field theory in the limit of infinite quartic self coupling.) The
discretization on a lattice replaces the derivative term by the nearest
neighbor differences, $\partial_\mu\phi(x) \longrightarrow
(\phi_{x+\hat{\mu}}-\phi_x)/{l_{x,x+\hat{\mu}}}$, where $l_{x,y}$
denotes the distance between the neighboring two points.  The area
differential $d^2x$ has to be replaced by an area element
$A_{x,\hat{\mu}}$ assigned to each lattice link (or site term).  The
discretized kinetic term then leads to the link term of the lattice
action (\ref{ActionDef}), with the weight factor
$w_{x,\hat{\mu}}=  2 A_{x,\hat{\mu}} /l_{x,x+\hat{\mu}}^2$.

\begin{figure}
\begin{center}
\epsfig{file=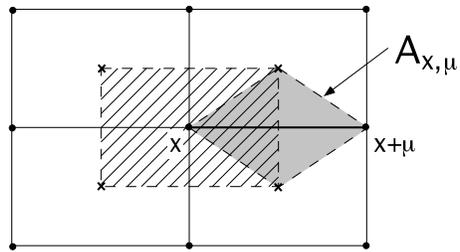,width=6cm}
\end{center}
\caption{\label{DualWeightElement}
The rectangular hatched area as attributed to the site $x$ is defined
by the dual sites (indicated by crosses) and is proportional to the
shaded area $A_{x,\hat{\mu}}$ that corresponds to the link from $x$ to
$x+\mu$.}
\end{figure}

The total lattice volume has to be split into lattice area elements
attributed to the link contributions.  These lattice area elements are
constructed with help of the dual lattice, where each site may be
defined as the barycenter of the plaquettes of the original lattice
\footnote{In \cite{ChFrLe82} the center of periphery was used; there is
no such point in a general quadrangle, therefore the center of mass
seems to be a reasonable alternative.} (cf. fig.
\ref{DualWeightElement}).  Links of the dual lattice are drawn between
sites corresponding to neighboring plaquettes. Each link on the
original lattice corresponds exactly to one link of the dual lattice.

For simplicity we make two approximations:
\begin{itemize}
\item We assume that all quadrangles are flat; in reality the four
corners are not co--planar. In the thermodynamic limit the lattice becomes
locally flat and the error vanishes.
\item We neglect angular distortions and approximate each quadrangle by
a rectangle; this allows us to write the area assigned to a lattice
link $x$ as the product of the lengths of the dual link
with that of the link itself, $A_{x,\hat{\mu}} = \frac{1}{2}\;
l'_{x,\hat{\mu}}\times l_{x,\hat{\mu}}$. The error introduced due to
this approximation does not vanish in the thermodynamic limit and
introduces slight distortions from the regular spherical surface.
However, the curvature is still smeared out over the whole lattice,
although somewhat non--uniformly. As will turn out later, that appears
to be no serious problem for the finite size analysis.
\end{itemize}
This way the total lattice volume is distributed over all links.

With these approximations we find that the weight factors for the
lattice geometry $S[N]$ are just the ratio of dual link length to the
link length
\begin{equation}
w_{x,\hat{\mu}} \equiv l_{x,\hat{\mu}}' / l_{x,\hat{\mu}}
\end{equation}
equivalent to the factor suggested in \cite{ChFrLe82} for triangulations.

For this particular sphere--like lattice we do not expect, that the
thermodynamic limit value of the critical coupling is identical to
that of the torus.
Since we have introduced a change of the action affecting all links,
$\beta_c$ will be renormalized. Still we expect universal behavior of
the critical exponents, to be checked numerically.

The computer programs use index tables to deal with the lattice
geometry.  For the lattices $S[N]$ the weight factors are precalculated
and tabulated. In our discussions we will refer to the lattice volume
\begin{equation}
V=\sum_{x,\mu}w_{x,\hat{\mu}}
\end{equation}
as the typical size quantity. For lattices ${T\!O\,}$,
${P\!I\,}$ and ${S\!H\,}$
this is just the number of links. A length scale may be defined as
$L\equiv\sqrt{V/2}$, such that its value is $N$ for a torus
${T\!O\,}[N]$.

In another study of the Ising model \cite{DiGoSa94} regular honeycomb
lattices folded to tetrahedron shapes have been used. There it was
possible to obtain series expansions as well as Monte Carlo results.
In \cite{HoJa95} the Ising model was studied with Monte Carlo methods
on triangulated random lattices with sphere--like topology.  We compare
these results with ours in the section \ref{sec4}.

\section{Simulation details\label{sec3}}

In the Monte Carlo simulation we studied lattices with the base length
$N=$ 16, 32, 64 and 128. (The torus results were determined up to
$N=256$.) The lattices have the volumes given in table \ref{VolTable}.

\begin{table}[ht]
\begin{center}
\begin{tabular}{|c|r|r|r|r|}
\hline
Base Length $N$
&$T\!O\,$&$P\!I\,$&$S\!H\,$&$S$\\
\hline
\hline
16&512&1020&2700&2724.8\\
32&2048&4092&11532&11650.0\\
64&8192&16380&47628&48147.0\\
128&32768&65532&193548&195726.9\\
\hline
\end{tabular}
\caption{The volumes $V$ as defined in the text; the equivalent linear
extent $L=(V/2)^{\frac{1}{2}}$ ranges from 16 (for the torus) up to 313
(for the sphere).}
\label{VolTable}
\end{center}
\end{table}

For the updating of the spins we used the Swendsen--Wang cluster
algorithm \cite{SwWa87}.  For each lattice size the bulk energy
histograms were determined for various values of the coupling and then
combined with help of the Ferrenberg--Swendsen multihistogram technique
\cite{FeSw}.  This leads to an optimal estimator for
the distribution densities $\rho_L(E)$ for the partition function
\begin{equation}\label{PartFcn}
Z_L(\beta) =\sum_E \rho_L(E) \exp(-\beta E)
\end{equation}
and allows to determine various moments $\langle f(E) \rangle$.  Of
course the statistical accuracy deteriorates when one evaluates these
quantities outside the domain of $\beta$--values, where one determined
the histograms. Suitable overlap between the individual histograms is
necessary for a reliable application of the method.  We determined only
the energy histograms and analyzed only observables in the even sector
of the model and not the observables that involve the magnetization.

We measured the specific heat, the Challa--Landau--Binder
cumulant \cite{ChLaBi86} and another $4^{\mbox{\footnotesize th}}$
order cumulant suggested by Binder (cf. the review \cite{Bi92}):
\begin{eqnarray}\label{cumdefs}
c_V(\beta,L)&=& ~~\frac{1}{V} \langle (E-\langle E\rangle )^2
\rangle \; ,\\
V_{CLB}(\beta,L)&=& -\frac{1}{3} \frac{\langle (E^2-\langle
E^2\rangle )^2\rangle}{\langle E^2\rangle^2} \; ,\\
U_4(\beta,L)&=& \frac{\langle (E-\langle E\rangle)^4\rangle}
{\langle(E-\langle E\rangle )^2\rangle^2}  \; .
\end{eqnarray}
The positions and values of their respective extrema are used for the
FSS analysis.

Eq. (\ref{PartFcn}) defines implicitly an analytic continuation to
complex values of $\beta$ not too far away from the real axis.
Therefore it is  possible to determine the nearby zeros of the
partition function \cite{YaLe52} in the complex $\beta$--plane, the
so--called Fisher zeros \cite{Fi68}.  As will be demonstrated below, in
particular the imaginary part of the zero closest to the real axis
provides a high quality estimator for the critical exponent $\nu$
with small corrections to the leading FSS behavior (cf.
\cite{KeLa} for a recent high statistics study of the Ising model in
4D, where it was possible to identify the logarithmic corrections to
scaling on basis of the Lee--Yang singularities \cite{YaLe52}). So  real
and imaginary parts of the closest Fisher zeros provide further (even)
observables.

For each lattice size we simulated the system at up to 20 different
values of $\beta$ between 0.41 and 0.47. The integrated autocorrelation
times for the energy varied between $\tau\approx 3-7$. For each
size we produced between 2 and $5 \times 10^6$  independent configurations.
The errors were estimated with the jackknife algorithm, i.e. from the
variation of the results for the analysis of subsamples of the raw
data.

Throughout the discussion of the data
we use $L\equiv\sqrt{V/2}$. The fit quality
is expressed through the goodness of fit parameter $Q$.
\footnote{The goodness of fit parameter $Q$ is defined as
$ Q=\Gamma(\frac{n-p}{2},\chi^2)/\Gamma(\frac{n-p}{2}) $,
where $n$ is the number of fit points and $p$ is the number of fit
parameters. It is the integrated probability over all $\chi^2$ larger
than the measured one.}

\section{Results and analysis\label{sec4}}

\subsection{Finite size scaling and topology}

{}From the usual scaling hypothesis \cite{FiBaBrCa,Ba83} one expects
for the singular part of the free energy density the scaling
behavior
\begin{equation}
f(\tau,L) = L^{-1/D} f(\tau L^{1/\nu}, 1)\;,
\end{equation}
where $\tau = (1-\beta/\beta_c)$ denotes the reduced coupling
and $L$ is the length scale.
{}From this one derives the scaling behavior of the cumulants.
At a second order phase transition we expect (for $D=2$)
\begin{eqnarray}
C_{max}(L)	&\simeq &
\left\{
\begin{array}{ll}
   L^{\alpha/\nu} & (\mbox{for~}\alpha>0)\\
   O(\ln L)	  & (\mbox{for~}\alpha=0)
\end{array}\right. \label{FSSrelations}\\
V_{CLB,min}(L) 	&\simeq &
\left\{
\begin{array}{ll}
L^{\alpha/\nu - 2} & (\mbox{for~}\alpha>0)\\
L^{-2} \ln L 	   &(\mbox{for~}\alpha=0)
\end{array}\right. \label{FSSV}\\
U_{4,min}(L)  	&\simeq	&
\left\{
\begin{array}{ll}
O(1) + O(L^{-\alpha/\nu}) &(\mbox{for~}\alpha>0)\\
O(1) + O(1/\ln L)^2 &(\mbox{for~}\alpha=0)
\end{array}\right. \label{FSSU4}\\
\mbox{Im}\;z_0(L)& \simeq & L^{-1/\nu} \label{FSSIm}\\
\beta_{c}(L)-\beta_c  &\simeq& L^{-\lambda} \label{FSSbeta}
\end{eqnarray}
with Josephson's law $\alpha = 2 - D\nu$.  The asymptotic
value of $U_4$ depends
on the details of the distribution density $\rho(E)$ and is e.g. 3 for
a Gaussian distribution. For the Ising model the Onsager solution
gives $\nu=1$ and $\alpha=0$.

We denote by $\beta_c(L)$ our definitions for pseudocritical points:
The positions of the extrema in the cumulants. The so--called
shift--exponent $\lambda$ is for
many models equal to $1/\nu$, but not necessarily so in general; this
relation is not a necessary conclusion of FSS (cf. the discussion in
\cite{Ba83}). We return to this issue in the discussion of the results.

A priori we know nothing about the absolute size of the multiplicative
coefficients in the scaling formulas. They depend on the details
of the lattice geometry and topology and on the boundary conditions.

The FSS behavior comes from the rescaling properties of the bulk
quantities. The effect of changing the boundary properties may be
responsible for further contributions. The non--homogeneous distribution
of the curvature in our lattices ${P\!I\,}$ and
${S\!H\,}$ might also be
responsible for additional (constant) terms in the total free energy.

Actually, since the total curvature is an invariant, there may be
another contribution, which --- relative to the bulk contribution
$O(V)$ --- becomes irrelevant in the thermodynamic limit.  It has been
shown \cite{CaPe88}, that the total free energy for finite 2D systems
with non--singular metric and smooth boundaries has at criticality (in
addition to boundary terms $O(L)$) an asymptotic contribution
proportional to $\ln L$. The proportionality constant is a product of
the central conformal charge and the Euler number (vanishing for the
torus). Thus this contribution depends only on the topology of the
system, not on the shape of the boundary.

All these contributions to the free energy are suppressed $O(1/L^2)$ or
$O(\ln{L} /L^2)$ relative to the leading term.  In the absence of a
strict theory we therefore might expect corresponding additive
corrections terms in the FSS relations (\ref{FSSrelations}) --
(\ref{FSSbeta}).

\subsection{Cumulant values and partition function zeros}

The values of the cumulants at their respective pseudocritical points
provide information on the critical exponents according to
(\ref{FSSrelations}) -- (\ref{FSSU4}). As discussed, they may have
geometry dependent corrections. However, in our data we find
qualitatively excellent agreement with the scaling of the torus--results
and no significant indication of geometry--corrections.

\begin{figure}
\begin{center}
\epsfig{file=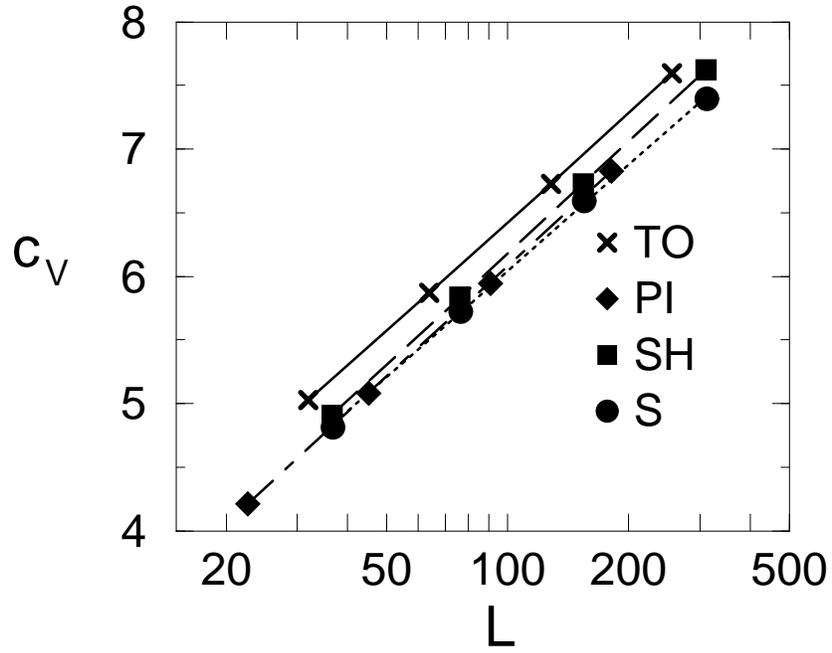,width=11cm}
\end{center}
\caption{\label{SpecHeat} Specific heat peak values vs. $\ln L$ for
lattices ${P\!I\,}$, ${S\!H\,}$, ${T\!O\,}$ and
$S$. Here and in all the other figures the error bars are smaller
than the symbols.}
\end{figure}

Fig. \ref{SpecHeat} shows that the specific heat scales with $\ln L$,
as expected for $\alpha =0$. A power law fit gives a value for $\alpha$
compatible with 0 and has a larger $\chi^2$: The logarithmic behavior
is preferred.

Comparing the results for the higher order cumulants $V_{CLB}$ and
$U_4$ we also find excellent agreement with the torus results, if
compared at the corresponding scales $L$, and with the expected scaling
behavior for $\alpha=0$ according to (\ref{FSSV}) -- (\ref{FSSU4}). In
particular the results for $V_{CLB}$ lie on top of a common curve for
all geometries.

\begin{figure}
\begin{center}
\epsfig{file=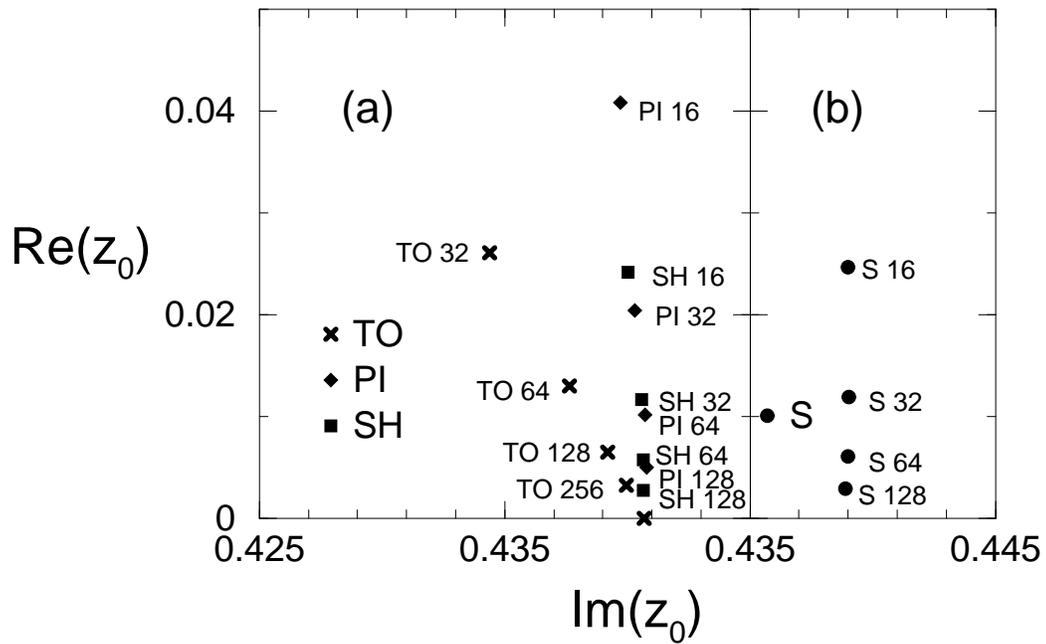,width=14cm}
\end{center}
\caption{\label{FisherZeros}(a) Position of the partition function zero
closest to the real axis in the complex $\beta$ plane for different
lattices.The numbers indicate the base length $N$.  The real part of
the zeros are closer to their thermodynamic value for the sphere--like
lattices than for the torus. In (b) we plot the results for the
spherical surface lattices $S$.  }
\end{figure}

The finite size dependence of the positions of the partition function
zeros confirms this observation.  In fig.  \ref{FisherZeros} the
nearest Fisher zeros for sphere--like lattices are compared with those
for toroidal lattice.  The real part is substantially closer to the
thermodynamic limit. Its scaling properties are discussed below
together with the pseudocritical points derived from the cumulants.

\begin{figure}
\begin{center}
\epsfig{file=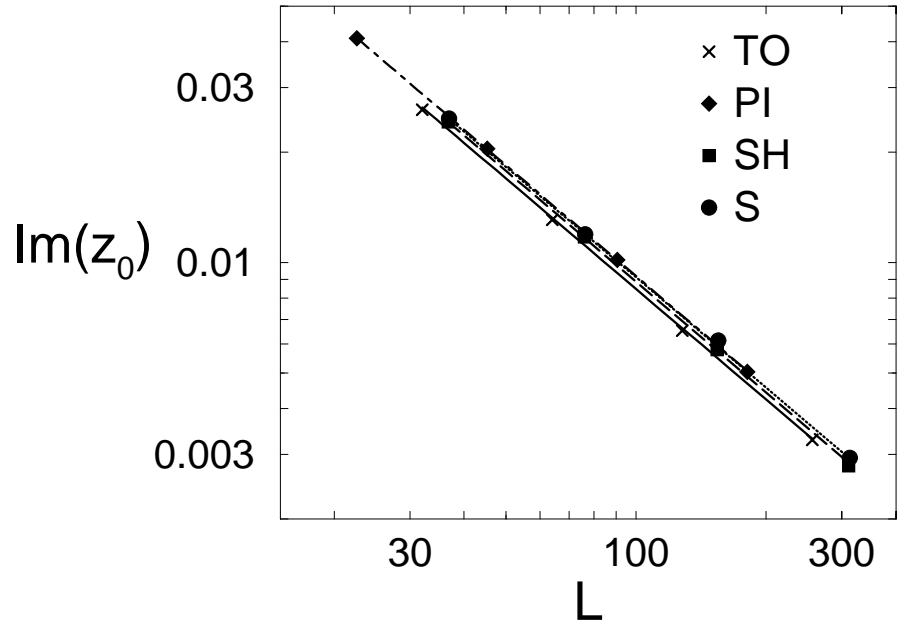,width=12cm}
\end{center}
\caption{Plot of $\ln (\mbox{Im}\;z_0)$ vs. $\ln L$; the fit
represents the leading FSS behavior.
\label{ImZeroLogLog}}
\end{figure}

The imaginary part of the closest Fisher zero appears to profit from
the smallness of the deviation of the real part from the thermodynamic
value. The log--log plot (fig. \ref{ImZeroLogLog}) demonstrates the
excellent scaling signal and a fit of the form
\begin{equation}
 \label{imfit}
\mbox{Im}\;z_0(L)=a L^{-1/\nu}
\end{equation}
gives $\nu=0.9964(46)$ (goodness of fit $Q=0.39$) for ${P\!I\,}$,
$\nu=0.9975(57)$ ($Q=0.77$) for ${S\!H\,}$ and $\nu=1.0023(54)$
($Q=0.12$) for $S$ lattices. For all these lattices the result
is in perfect agreement with the value $\nu=1$ of the toroidal lattice.

We conclude, that the imaginary part of the first partition function
zero is an optimal observable for extracting the critical
exponent $\nu$. It appears to be least affected by correction to scaling
due to lattice topology and boundary effects.

In this light the excellent scaling behavior of the specific heat
should not be too surprising, since the peak value is directly related
to the vicinity of the closest Fisher zero. Since the partition
function is proportional to the product of all zeros,
\begin{equation}
Z\propto \prod_i (\beta - z_i)
\end{equation}
the specific heat includes for its singular part the
contribution
\begin{equation}
\sum_i \frac{1}{(\beta - z_i)^2}\;.
\end{equation}
The closest zero therefore contributes a term $\propto (\mbox{Im}\;z_0)^{-2}$
to the peak value of $c_V$ \cite{KeLa}.

\begin{figure}
\begin{center}
\epsfig{file=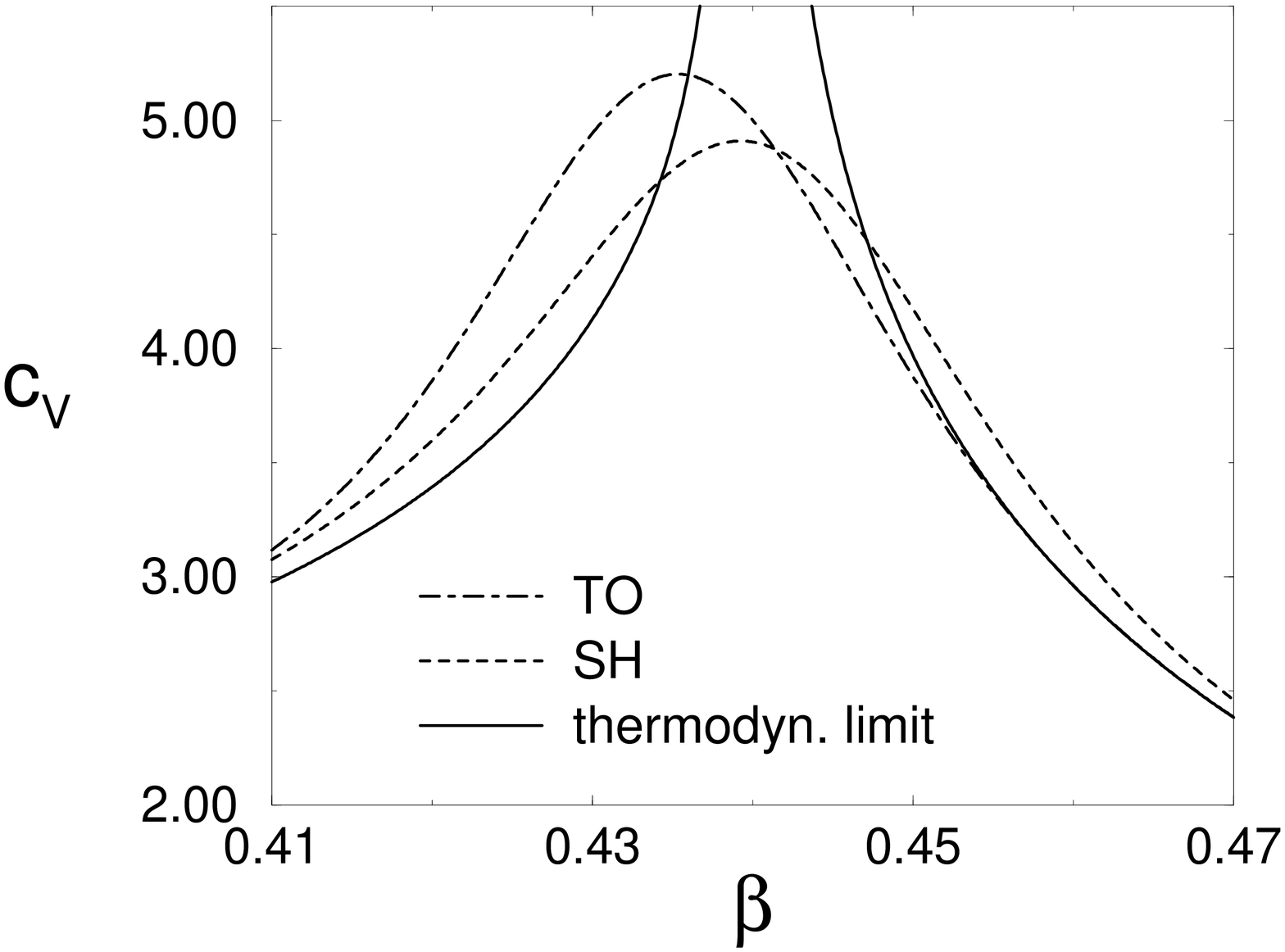,width=12cm}
\end{center}
\caption{\label{fco} Specific heat per unit volume vs. $\beta$ for
${T\!O\,}[37]$ and ${S\!H\,}[16]$; these lattices of similar volume are
compared to the Onsager solution.  One finds that the curve for the
${S\!H\,}$ lattice is more symmetric around $\beta_c$ than for the
${T\!O\,}$.}
\end{figure}

The fact, that the closest zero approaches the real axis (with
increasing lattice size) almost perpendicular is also clearly exhibited
by the shape of the specific heat itself. In fig. \ref{fco} we compare
the torus results with those for the cubic surface lattices for
equivalent lattice volumes.  The approach to the infinite volume case
(Onsager solution) is in a much more symmetric way than for the torus
lattice.

\subsection{Pseudocritical points}

We discuss here the pseudocritical values derived from
\begin{itemize}
\item[-] the peak positions of the specific heat,
\item[-] the minima positions of the other cumulants $V_{CLB}$ and $U_4$,
\item[-] the real part of the position of the closest
zero in the complex $\beta$--plane.
\end{itemize}
For the lattice geometries ${P\!I\,}$ and ${S\!H\,}$ we
expect (see the discussion
in sect.  \ref{sec2}) that in the thermodynamic limit the critical
values $\beta_c$ coincide with those of the torus, and we therefore
present these results in direct comparison.  For the
lattice type $S$ the asymptotic value  of the critical coupling will be
somewhat different and we  discuss these results separately.

\subsubsection{${P\!I\,}$ ~and ${S\!H\,}$ ~lattices}

It turns out, that both lattice geometries have very similar behavior
and agree (except for the smallest lattice ${P\!I\,}[16]$) even
numerically
with each other, if compared at corresponding volumes.

\begin{figure}
\begin{center}
\epsfig{file=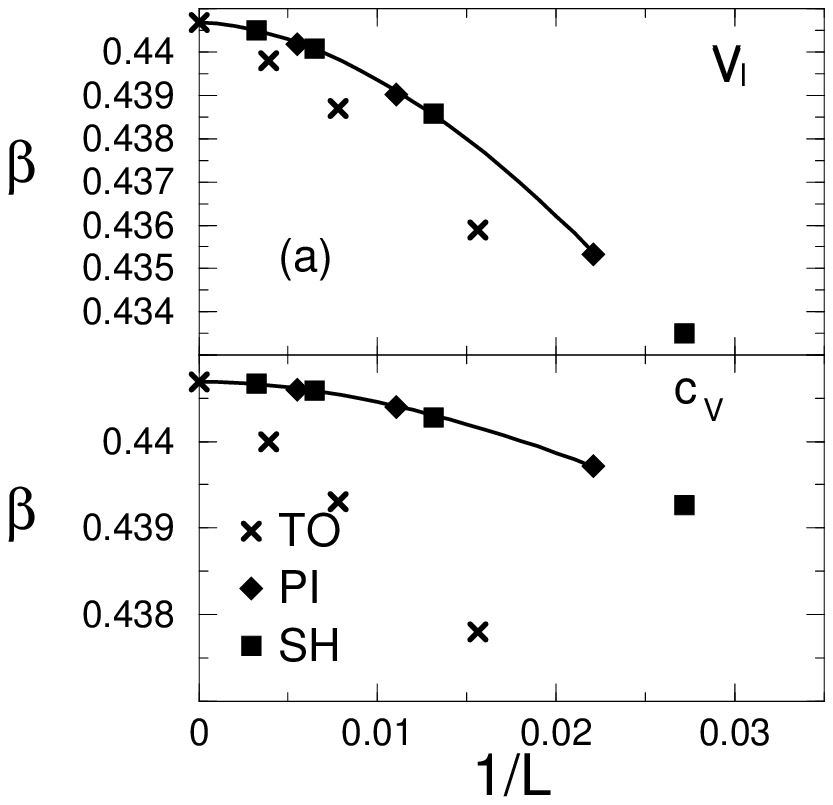,width=8cm}
\epsfig{file=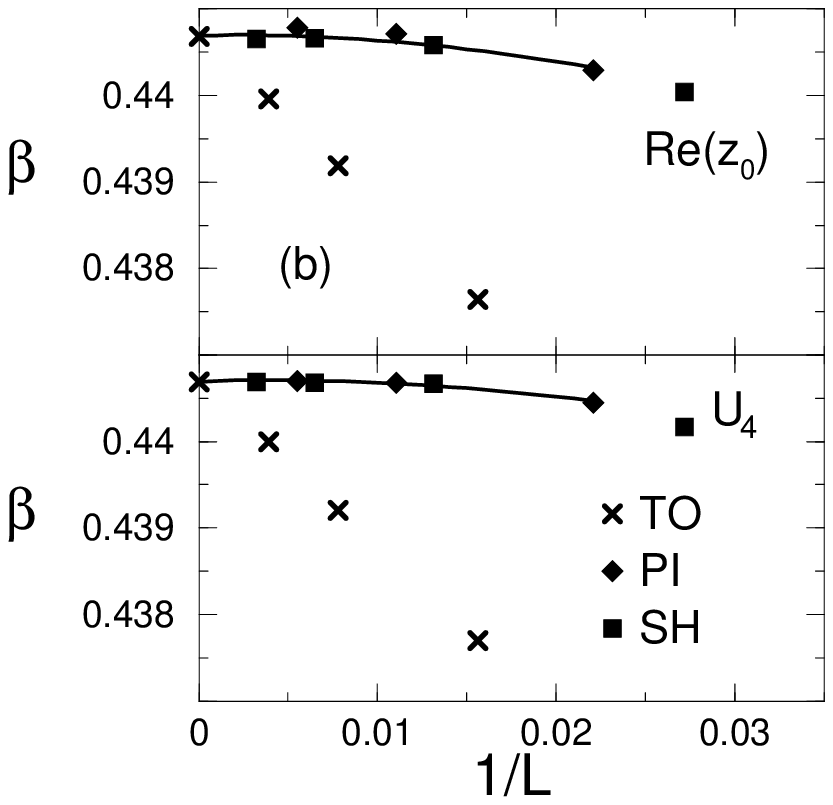,width=8cm}
\end{center}
\caption{ \label{FigExtrema} Results for the pseudocritical coupling
vs. $L$ for (a) the specific heat $c_V$ and $V_{CLB}$, (b) the real
part of the closest partition function zero and $U_4$, for pillow, dual
cube and torus type lattices.  The curves are fits according to
(\ref{betacFITSa}) for the pillow and cubic surface lattices as
described in the text. }
\end{figure}

As fig. \ref{FigExtrema} clearly exhibits, there is an obvious
difference in the FSS behavior compared to the usual torus results.
The overall size of the corrections to the thermodynamic value of the
critical coupling are much smaller for the sphere--like lattices.  The
leading FSS behavior of $\beta_c(L)$ for large $L$ should follow
(\ref{FSSbeta}). For the  Ising model on a torus the shift exponent is
$\lambda=1/\nu =1$ \cite{FeFi69}.  This leading behavior linear in
$1/L$ is evident in the figure.  However, for ${P\!I\,}$ and ${S\!H\,}$
another effect seems to blur this picture: A possible (but clearly very
small) linear term is dominated by contributions nonlinear in $1/L$.

As discussed in \cite{Ba83} the leading linear term may vanish even for
$N\times M$ torus geometry, depending on the ratio $N/M$.  In
particular it vanishes in the limit $M\to \infty$, where the leading
behavior becomes $O(\ln L /L^2)$ \cite{FeFi69}. There are also other
specific models and cases, where $\lambda\neq 1/\nu$ \cite{Ba83}.  As
mentioned below (\ref{FSSbeta}) also the topology may give rise to
additional terms \cite{CaPe88} in the free energy (per unit volume)
proportional to the Euler number and to $O(\ln L /L^2)$; it is unclear
how these affect the pseudocritical points in our particular
situation.

This observation, that the dominating behavior appears to be non--linear
in $1/L$, was also made in a study of the Ising model for a honeycomb
lattice on a tetrahedron surface \cite{DiGoSa94}. Both, series and Monte
Carlo results led to a value $\lambda=1.745(15)$ \cite{DiGoSa94}.
The value of $\nu$ obtained there from the correlation length and the
specific heat agreed with the Onsager value.

We therefore fit our data for the pseudocritical points to
\begin{eqnarray}
\beta_c(L)-\beta_c&=&a L^{-1}+b L^{-\lambda} \quad ,\label{betacFITSa}\\
\beta_c(L)-\beta_c&=&a L^{-1}+b L^{-2}\ln{L} \quad ,\label{betacFITSb}
\end{eqnarray}
with the Onsager value for $\beta_c = \frac{1}{2}\ln{(1+\sqrt{2})}$.

The data for each of the 4 definitions of pseudocritical coupling
(from $c_V$, $V_{CLB}$, $U_4$ and $\mbox{Re}\; z_0$)
appears to be consistent for both geometries ${P\!I\,}$ and
${S\!H\,}$.  We therefore
use one set of parameters $a$ and $b$ different for each definition but
identical for the two geometries. The value of $\lambda$ is assumed to
be universal for all definitions and both geometries. We fit to data for
$N\geq 32$.  The results
according to (\ref{betacFITSa}) are given in table \ref{FITPars} and
are plotted in fig. \ref{FigExtrema}:  they fit the data perfectly with
a $\chi^2/d.f.\simeq 1.2$ $(Q=0.26)$.

\begin{table}[ht]
\begin{center}
\begin{tabular}{|l|l|l|r|r|r|r|}
\hline
Lattice&$\lambda$&Par.&$c_V$&$V_{CLB}$&$U_4$&Re $z_0$\\
\hline
\hline
${P\!I\,}$,${S\!H\,}$&1.76(7)	&$a$	&0.002(4)
&0.000(13)	&-0.011(5)	&0.009(10)\\
&		&$b$	&-0.83(18)	&-4.29(96)	&-0.37(13)
&-0.45(24)\\
\hline
$S$&	1.71(10)&$a$	&0.028(10)	&0.025(28)	&0.032(11)
&0.029(11)\\
&		&$b$	&-0.62(13)	&-3.34(86)	&-0.24(14)
&-0.29(13)\\
\hline
\end{tabular}
\end{center}
\caption{\label{FITPars} Scaling law coefficients $a$, $b$ and exponent
$\lambda$
for the various lattice geometries. Fits are according to (\ref{betacFITSa})
to data for $N=32 \ldots 128$ for ${P\!I\,}$ and  ${S\!H\,}$
and $N=16 \ldots 128$ for $S$}
\end{table}

As expected from looking at the data we find only a small contribution
to the term $O(1/L)$, compatible with zero for all observable except for
$U_4$. Removing this term altogether seems conceivable,
although $\chi^2$ is quadrupled in this case. The second
term clearly dominates. However, the resulting value
$\lambda=1.76(7)$, although consistent with the results for the
tetrahedron \cite{DiGoSa94}, is not stringent. In fact, allowing for the
second ansatz (\ref{betacFITSb}) give almost the same fit quality and
would be {\em indistinguishable in the figure}. Also fixing $\lambda$ to a
value $2$ is still compatible with the data.

We conclude that we are in a situation where a possible leading FSS
term $O(1/L)$ has an (almost or completely) vanishing coefficient and
the subleading terms dominate. This resembles the Ising
model on a cylinder with infinite extension in one direction. It cannot
be decided, whether the corresponding term is of form
(\ref{betacFITSa}) or (\ref{betacFITSb}).

\subsubsection{Spherical surface lattices}

Now we turn to the approximate spherical surface topology ($S$).
Whereas for the pillow and cubic surface lattices the curvature is
concentrated in 8 or 24 points here it is more or less uniformly
distributed among all sites of the lattice. The total curvature (the
Euler number) remains constant.

\begin{figure}
\begin{center}
\epsfig{file=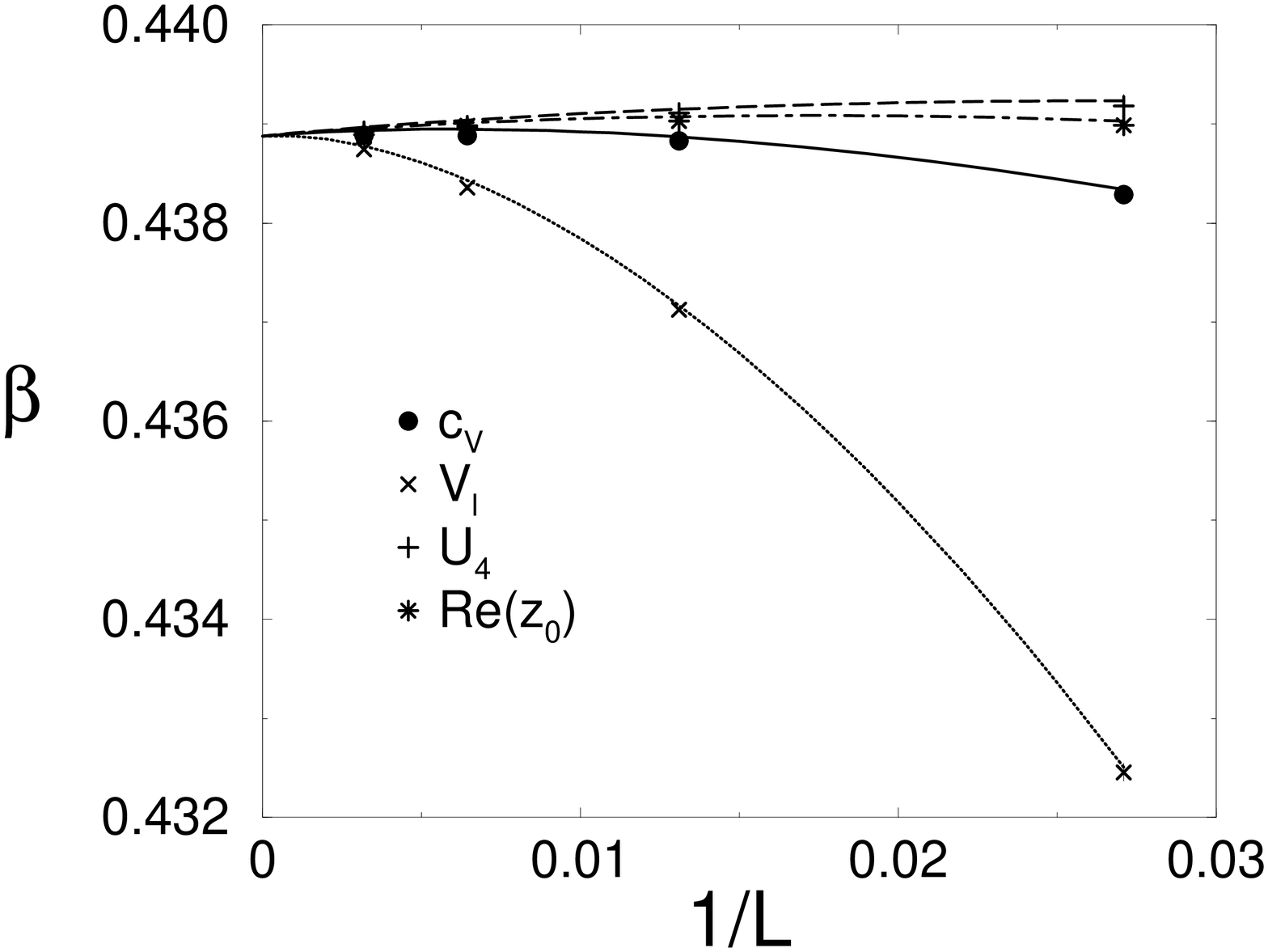,width=12cm}
\end{center}
\caption{\label{f10} Peak positions of the cumulants on spherical
surface lattices $S$. One can clearly see, that $\beta_c$ differs from
the critical temperature on the torus $\beta_{c,TO}\approx 0.44068$.
The curves are from fits to (\ref{betacFITSa}.) The error bars are
smaller than the symbols.}
\end{figure}

In fig. \ref{f10} the peak positions of the specific heat and the other
cumulants are plotted  together with fitted curves according to
(\ref{betacFITSa}). Here $\beta_c$ is a free parameter, otherwise we
follow the procedure discussed above, i.e.  one common value of
$\lambda$ but different parameters $a$ and $b$ depending on the
observable. Since we have fewer data we include the data from the
smaller lattices with $N=16$. This is also justified by the overall
smaller deviations from the asymptotic value.

The resulting values are also given in table \ref{FITPars}. We find a
behavior in agreement with  the other sphere--like lattices. The
contribution of the term $O(1/L)$ is again very small and the
non--linear term dominates again.  The overall variation of the
pseudocritical points with $L$ is for most observables smaller than for
the other lattice geometries.  The fitted critical temperature is
$\beta_c=0.43883(3)$ with a $\chi^2/d.f.\simeq 0.5$ ($Q=0.81$).  The
value of $\lambda=1.71(10)$ is consistent.

Again the fit to (\ref{betacFITSb}) gives results of comparable quality
and corresponding curves would be indistinguishable in the figure.

\section{Conclusion\label{sec5}}

We have performed a high statistics Monte Carlo study of the Ising
model on lattices of various size and different shapes, all with
sphere--like topology.  Our FSS analysis led to the following
conclusions.
\begin{itemize}
\item
We have found explicitly, that the Ising model on a spherical surface
topology lies in the same universality class as the planar Ising model
with periodic boundary conditions --- topologically the surface of a
torus.  Our results demonstrate that universality holds, independent of
the lattice geometry. This agrees with
similar conclusions obtained for tetrahedral lattices \cite{DiGoSa94}
and random lattices \cite{HoJa95} of sphere--like topology.
\item
However, some observables are not well suited to
find the expected leading FSS behavior.
Different observables vary in their sensitivity. Of the studied
quantities (in the even sector of the Ising model) we find that the
imaginary part of the Fisher zero closest to the real axis has the
smallest (in fact: not identifiable within our accuracy) deviations
from the leading FSS behavior. Related to this quantity, also the peak
value of the specific heat scales according to the FSS formulas with
the Onsager values for the critical exponents without further
(identifiable) corrections. The values of the other cumulant have
larger statistical errors but are also in agreement with the torus
results.
\item
The change in the topology class influences the size of the FSS
contributions. This appears to affect in particular the pseudocritical
points. We find no significant contribution of $O(1/L)$, which is the
dominant FSS term for the torus pseudocritical points (with a shift
exponent $\lambda=1/\nu=1$).  Instead we find that the FSS behavior is
dominated by a term $O(L^{-\lambda})$ with a mean value $\lambda\simeq
1.74(6)$ (averaging the results for ${P\!I\,}$, ${S\!H\,}$
and $S$). A compatible
value was obtained in an independent study for tetrahedral lattices
\cite{DiGoSa94}.  The behavior seems to be universal for all
sphere--like lattices, independent of the details of the geometry.  This
contribution is also consistent with a term $O(\ln L/L^2)$; such a term
describes the FSS of the specific heat peak position for the Ising
model in cylinder geometry \cite{FeFi69}.  It also has been argued, that
a term of that kind contributes to the free energy per unit volume for
systems with non--zero Euler number \cite{CaPe88}.
Unfortunately this implies that none of these observables is
qualified to derive the critical exponent $\nu$.
\item
In general we find that the studied sphere--like lattices have smaller
corrections to the infinite volume behavior than one observes for the
torus (i.e. periodic boundary conditions). The approach to the
thermodynamic shapes is faster and in a more symmetric way.
\item Our analysis
of the Fisher zeros of the partition function is consistent with this
picture.  With increasing size the closest zero approaches the real axis
almost perpendicular. (Note, that this
behavior is a finite size behavior and is not identical to the
asymptotic impact angle -- defined e.g. as the angle between the first
and second zero).  The results for the spherical surface geometry
appear to be closest to the thermodynamic behavior in general.  For
different lattice (sphere--like) geometries the FSS behavior is
consistent if one chooses the size variable $L=\sqrt{V/2}$, where $V$
is the number of links. This choice appears to be preferable over the
base length $N$.
\end{itemize}

In general our conclusion are consistent with other results on sphere--like
lattices for the Ising model \cite{DiGoSa94,HoJa95} and with similar
observations in other models \cite{HoJaJe95}, also in higher dimensional
lattices \cite{JeLaNe95}.

How can one explain the more symmetric and seemingly faster approach to
the thermodynamic limit, that occurs for the sphere--like lattices as
compared to the torus? One may argue, that on the 2D torus there are
two globally distinguished directions.  As soon as the correlation
length $\xi$ approaches some fraction of the linear size, the
system notices the loss of its rotational invariance. Larger clusters
may then span in the two distinguished directions of the lattice.  On
sphere--like surfaces (although there is local orientation) there are no
globally distinguished directions. Rotational invariance holds to a
larger extent and the behavior of the finite system is more symmetric
around the critical point.

\subsection*{Acknowledgment}

We are most grateful to J. Jers\'ak, M. L\"uscher and T. Neuhaus for
many helpful suggestions and discussions.  We appreciate stimulating
conversations with H. Gausterer, A. Jakovac and W. Janke.

\newpage

\end{document}